\documentclass[aps,prl,twocolumn,floatfix]{revtex4}
\usepackage{graphics,epsf,bm}

\def\w{\omega}

\def\bk{{\bf k}}

\def\D2n{\Delta^2(s, \w_n)}
\def\enk{\epsilon({\bf k})}
\def\ek{\epsilon_{\bf k}}
\def\dk{\delta_{\bf k}}
\def\enQ{\epsilon({{\bf k}+{\bf Q}})}
\def\cka{c_{{\bf k},\alpha}}
\def\cdka{c^\dagger_{{\bf k},\alpha}}
\def\apdka{a^\dagger_{+1,{\bf k},\alpha}}

\def\amdka{a^\dagger_{-1,{\bf k},\alpha}}

\def\ckqb{c_{{\bf k}+{\bf q} ,\beta}}
\def\cdkQa{c^\dagger_{{\bf k}+{\bf Q} ,\alpha}}

\def\uk{u_{\bf k}}
\def\uq{u_{{\bf k}+{\bf q}}}
\def\vq{v_{{\bf k}+{\bf q}}}
\def\vk{v_{\bf k}}

\def\Sq{{\bf S}_{\bf q}}
\def\Smq{{\bf S}_{-{\bf q}}}

\begin{document}

\title{Non-fermi liquid behavior in itinerant antiferromagnets}
\author{I.~Vekhter}
\affiliation{Theoretical Division, Los Alamos National Laboratory,
Los Alamos, NM 87545} \altaffiliation[Present address:
]{Department of Physics and Astronomy, Louisiana State University,
Baton Rouge, LA 70803}
\author{A.~V.~Chubukov}
\affiliation{Department of Physics, University of Wisconsin,
Madison, WI 53706}
\begin{abstract}
We consider a two dimensional itinerant antiferromagnet
  near a quantum critical point.
 We show that, contrary to conventional wisdom,
 fermionic excitations  in the ordered state
are not the usual Fermi liquid quasiparticles.
 Instead, down to very low frequencies,
the fermionic self energy varies as $\omega^{2/3}$.
This non-Fermi liquid behavior originates in the coupling of
fermions to the longitudinal spin susceptibility
 $\chi_{\parallel} (q, \Omega)$ in which the
 order-induced ``gap'' in the spectrum at $q=0$ dissolves into the
 Landau damping term  at $v_F q >\Omega$.  The transverse
spin fluctuations obey $z=1$ scaling characteristic of spin waves,
but remain overdamped in a finite range near the critical point.
\end{abstract}

\date{\today}

\maketitle

Fermi liquid (FL) theory is a powerful tool to study properties of
interacting electrons.  It states that, upon switching on the
electron-electron interaction, elementary excitations near the
Fermi surface remain fermionic, with well-defined energy and
momentum. As a result, Fermi liquids have a constant uniform spin
susceptibility and a linear in temperature, $T$, electronic
specific heat.

In the last two decades, many new compounds were found which
exhibit thermodynamic and transport properties dramatically
different from those of a Fermi liquid~\cite{piers_review}.
Understanding  the origin
of this non-Fermi liquid (NFL) behavior is one of the most
important challenges in condensed matter physics.

Few avenues leading to the NFL behavior in dimension $D>1$ have
been proposed in the theory of interacting electron systems. One
of the most widely studied scenarios ties the destruction of the
Fermi liquid to a quantum critical point (QCP) \cite{Sachdev}. At
a QCP the fluctuations of the order parameter(OP) are gapless,
 and the effective electron-electron interaction mediated
by this bosonic mode is long-ranged. Scattering of electrons by
gapless bosons is singular and destroys a Fermi liquid when the
local static bosonic susceptibility $\chi_l = \int d^{D-1} q~ \chi
(q, \omega=0)$ diverges ~\cite{acs}.

It is generally expected that, away from the QCP, the Fermi liquid
behavior is preserved.  On the disordered side the FL is protected
by the gap in the OP fluctuations due to a finite correlation
length. On the ordered side, where the OP acquires a mean value,
the amplitude (longitudinal) bosonic excitations are gapped, while
the gapless Goldstone excitations (phase, or transverse modes) are
harmless for fermions due to the Adler's principle~\cite{Adler}
that states that the fermion-boson vertex vanishes at the momentum
transfer equal to the ordering wave vector ${\bf Q}$.

The main conclusion of this Letter is that, contrary to this
general belief, a novel non-Fermi liquid electron behavior emerges
on the ordered side of a QCP when ${\bf Q}\neq 0$. This NFL
behavior originates from the interaction between electrons and
longitudinal bosonic excitations. The key observation is that this
mode becomes gapless and overdamped in the range $v_F q >\Omega$
due to Landau damping from the electrons, and  gives rise to a
strong electron-electron interaction even at the lowest energies.
The NFL behavior persists down to a frequency $\omega_{min}$ that
vanishes if the fermionic bandwidth is infinite.

Below we consider a SDW transition between a paramagnet and an
itinerant antiferromagnet (AFM)  with ${\bf Q} = (\pi,\pi)$ and
the dynamical exponent $z=2$ in $D=2$. Such a transition is both
one of the most studied and the most relevant experimentally, as
it occurs in many heavy fermion materials and is believed to be
responsible for their unusual properties.

We first present the results and discuss the physics and then
report the details  of the  calculations. We measure the AFM order
by the gap in the fermionic spectrum, $m$. Opening of this gap
renormalizes the spin susceptibilities from their form on the
paramagnetic side, $\chi^{-1}_\perp ({\bf q},\Omega)=\chi^{-1}_\|
({\bf q},\Omega)=({\bf q}-{\bf Q})^2 -i\gamma|\Omega|$ \cite{acs}.
Hereafter {\bf q} denotes the deviation from $(\pi,\pi)$, except
when noted. Conventional wisdom holds that, in the AFM state, at
energies smaller than $2m$, the transverse spin excitations are
the Goldstone spin waves, while the longitudinal excitations are
gapped at all bosonic ${\bf q}$ and $\Omega$.  We find, however,
that this behavior holds only when $v_F q$ is smaller than the
frequency, $2m >\Omega> v_F q$. In this range, $\chi^{-1}_{\perp}
\approx q^2 -a\gamma\Omega^2/m$, has a spin-wave form
 ($a$ is a number $O(1)$),
 while $\chi^{-1}_{\parallel} \approx q^2 + a\gamma
\sqrt{m^2-\Omega^2/4}$ has a gap, $2m$, for the propagating spin
excitations.

In the opposite limit,  $2m > v_F q > \Omega$, we find
\begin{equation}
\chi_{\parallel}^{-1} \approx
 q^2 - 2i \gamma m |\Omega|/v_F q,~~
\chi_{\perp}^{-1} \approx q^2 -i \gamma |\Omega| v_F q/m.
\label{new2}
\end{equation}
The transverse spin excitations become overdamped, though they
still show $z=1$ scaling. More importantly, the constant $2m$ gap
in the longitudinal spin excitations dissolves into the Landau
damping term, and the longitudinal excitations become {\it
gapless} with the dynamical exponent $z=3$. This implies that at a
given $v_F q \ll m$, there are two regions of $\Omega$ ($\Omega <
v_F q$ and $\Omega > 2m -v_F q$), where $\chi_{\parallel}^{\prime
\prime} (q, \Omega)$ is nonzero (see Fig.~\ref{fig3}).

The gapless overdamped form of the longitudinal susceptibility
strongly affects fermionic self-energy as the interaction vertex
for $\chi_\|$ is not reduced by Adler's symmetry principle.
Critical theories with $z=3$ have been studied extensively
~\cite{qc3}. In two dimensions, the electron self energy due to
interaction with such a mode varies as $(i\omega)^{2/3}$, i.e.
{\it the Fermi-liquid behavior is violated despite the presence of
the AFM order}. This self-energy comes from
 $v_F q \propto \Omega^{1/3} \gg \Omega$ where Eq. (\ref{new2}) is valid.
 Similar power law for the self energy in the ordered state
was argued  to be present in electronic nematic phases
\cite{Oganesyan}.

Above results are for an infinite fermionic bandwidth, $W$. When
$W$ is finite, the longitudinal spin fluctuations have a gap of
the order of $m^2/W$ even for $\Omega < v_F q$. This scale sets
the lower cutoff for $\omega^{2/3}$ behavior; the Fermi liquid
behavior is restored at lower energies. It is essential that in
itinerant AFM,  $m\ll W$ (see below), and the cutoff scale is
parametrically smaller than $m$.

\begin{figure}
    \label{fig1}
    \epsfxsize=3.5in \epsfbox{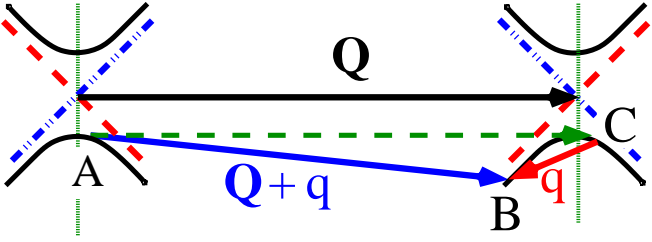}
    \caption{Doubling of the unit cell in an AFM transition. Vertical dotted
    lines are the boundaries of the magnetic Brillouin Zone. The Fermi surface
    at the ``hot spots'' at the paramagnetic side of the transition is the dashed
    red line, the dash-dotted blue line is its translation by {\bf Q}. Solid black
    lines denote the Fermi surface in the AFM state.}
\end{figure}

The  evolution of the spin response with $\Omega/v_F q$ is related
to the fact that the mass term in $\chi_{\parallel} (q, \Omega)$
is proportional to the particle-hole polarization bubble
${\widetilde\Pi} (q, \Omega) = \Pi (q, \Omega) - \Pi (q, 0)$, see
below. In a nested antiferromagnet, the AFM order opens a gap over
the entire Fermi surface. In this case, ${\widetilde\Pi} ({\bf q},
\Omega)$ is a constant independent on the ratio $\Omega/v_F q$.
However, in an itinerant AFM without nesting, only parts of the
Fermi surface close to the ``hot spots'' become gapped due to
long-range spin ordering. Elsewhere on the Fermi surface a
continuous particle-hole spectrum still exists. As a result, spin
excitations can decay into particle-hole pairs. On the
paramagnetic side, this process involves large momenta near ${\bf
Q}$, hence the damping is linear in frequency, $ {\rm Im }
{\widetilde\Pi}({\bf q}, \Omega)=
 \gamma |\Omega |$. In
the ordered state, the lattice period is doubled, and the
fermionic momenta ${\bf k}$ and ${\bf k}+ {\bf Q}$ become
equivalent. As a result, a bosonic mode at the momentum ${\bf Q}_1
\sim {\bf Q}$ scatters a fermion from a point ${\bf k}_F$  into a
Fermi surface point ${\bf k}_F+ {\bf q}$ with small ${\bf q} =
{\bf Q}_1 - {\bf Q}$, see Fig. \ref{fig1}. The polarization bubble
then becomes ${\widetilde\Pi} ({\bf q}, \Omega) \propto
\gamma m \Omega/\sqrt{\Omega^2 - (v_F q)^2}$. It gives rise to a
mass term for $\Omega \gg v_F q$, but reduces to $i \gamma m
\Omega/v_F q$ in the limit $\Omega \ll v_F q$.

We now provide the details of the calculation. To investigate the
properties of an itinerant antiferromagnet near a QCP we employ
the spin fermion model, which has been widely used to study the
properties of paramagnetic metals close to a magnetic instability
\cite{acs}. Here we use it on the ordered side of the transition.
The model Hamiltonian consists of three parts: the electronic
band, $H^{(0)}_{f}= \sum_{\bk, \alpha} \enk \cdka \cka$, the
collective fermionic spin excitations, $H^{(0)}_s= \sum_{\bf q}
\chi_0^{-1}({\bf q}) \Sq\Smq$, and the interaction of electrons
with the spin fluctuations,
\begin{equation}
    H^{(0)}_{sf}=g \sum_{{\bf k,q}, \alpha,\beta} \cdka
        {\bm{\sigma}}_{\alpha\beta} \Sq \ckqb + {\mathrm{h.
        c.}}.
    \label{HMain}
\end{equation}
Here $\enk$ is the bare quasiparticle dispersion, $\chi_0({\bf
q})$ is the bare static spin susceptibility
 and $g$ is a coupling
constant that we take to be independent of momentum. The rationale
for this model is that the antiferromagnetic interactions emerge
from the energies up to the fermion bandwidth. Therefore to
analyze the behavior of low-energy quasiparticles it is sufficient
to separate the energy scales and treat the collective spin
degrees of freedom as a separate bosonic field. This procedure is
internally consistent if the coupling $g\ll W$.

Properties of the spin subsystem are determined by the bare
susceptibility, which we take to have the Ornstein-Zernike form,
$\chi_0 ({\bf q})=\chi_0 /(\xi_0^{-2}+ ({\bf q}-{\bf Q})^2)$, as
in earlier work \cite{acs}. In the paramagnetic phase, the
magnetic correlation length, $\xi_0$, is real and controls the
separation from the AFM instability. The full dynamical
susceptibility is $\chi ({\bf q}, \w)=\chi_0/(\xi_0^{-2}+ ({\bf
q}-{\bf Q})^2 + {\widetilde\Pi}({\bf q}, \w))$, where ${\widetilde
\Pi}({\bf q}, \w)$ is the
full polarization bubble  that has to be evaluated within the
low-energy theory.

On the antiferromagnetic side of the transition $\xi_0$ becomes
imaginary ($\xi_0^{-2}<0$), so that the susceptibility diverges near
the AFM wave vector, {\bf Q}, leading to a staggered magnetic
moment, $\langle S^z_{\bf Q} \rangle$.
The divergence is compensated by $\Pi({\bf q}, 0)$
which acquires a finite value in the ordered phase. Furthermore,
for a finite $\langle S^z_{\bf Q} \rangle$,
the longitudinal ($zz$) and the transverse
polarization bubbles differ, leading to the anisotropic
susceptibility. The staggered moment, $\langle S^z_{\bf Q}
\rangle$, is related to $\xi_0^{-2}<0$, via the Goldstone requirement
that the fully renormalized $\chi^{-1}_{\perp} (q \rightarrow 0,
0)=0$.

Following the standard procedure, we introduce
 $m=g\langle S^z_{\bf Q} \rangle$,
include the ``condensed'' part of the spin fluctuations,
$\sum_{{\bf k},\alpha}
        \alpha m \cdkQa\cka$, into the
fermion part of the hamiltonian, and diagonalize it by  Bogoliubov
transformation to find $H_f=\sum_{n, {\bf k},
    \alpha} E_{n}({\bf k}) a^\dagger_{n, {\bf k},\alpha} a_{n, {\bf
        k},\alpha}$.
The dispersion of the fermions in the two new bands is given by
$E_{\pm 1} ({\bf k})= \ek\pm \sqrt{\dk^2+m^2}$ with $\ek=(\enk
+\enQ)/2$ and $\dk=(\enk - \enQ)/2$. The new and the original
fermion operators are related by a unitary transformation
    $\apdka=\uk\cdka -\alpha \vk \cdkQa$,
    and
    $\amdka=\alpha \vk \cdka +\uk \cdkQa$,
with $\uk^2 (\vk^2)=[1+(-)\dk/\sqrt{\dk^2+m^2}]/2$. The
interaction of fermions with the uncondensed part of the spin
excitations now takes the form
\begin{widetext}
\begin{eqnarray}
    \label{NewHsf}
   &&H_{sf}=g \sum_{n_1, n_2,{\bf k,q}, \alpha,\beta}
    \Gamma_{n_1,n_2}^{\alpha,\beta} ({\bf k},{\bf k}+{\bf q})
    a^\dagger_{n_1,{\bf k},\alpha} a_{n_2, {\bf k}+{\bf q},\beta}
    {\bm{\sigma}}_{\alpha\beta} {\bf S}_{{\bf q}-{\bf Q}} + {\mathrm{h.
        c.}}.,
\\
    \label{Vertex}
    &&\Gamma_{n_1,n_2}^{\alpha,\beta} ({\bf k},{\bf k}+{\bf q})=
    \biggl[\uk\uq-\alpha\beta\vk\vq\biggr]\biggl[1-\delta_{n_1,n_2}\biggr]
    -n_1\delta_{n_1,n_2}\biggl[\alpha\vk\uq+\beta\uk\vq\biggr].
\end{eqnarray}
The vertex for the Goldstone transverse mode,
$\Gamma_{n,n}^{\alpha,\bar\alpha} ({\bf k},{\bf k}+{\bf q})\propto
|{\bf q}|$ at small {\bf q}, as required by Adler's principle. The
 spin polarization bubble $\Pi (q, \Omega)$ is given by
    \begin{eqnarray}
    \Pi_{ij} ({\bf q}, \Omega) &=&
     \chi_0 T\sum_{n_1,n_2 \atop {{\bf k}\omega_n  \atop
     \alpha,\beta}}
    \sigma^{(i)}_{\alpha\beta} \sigma^{(j)}_{\beta\alpha}
    \Gamma_{n_1,n_2}^{\alpha,\beta} ({\bf k}+ \frac{{\bf q}}{2},{\bf k}- \frac{{\bf q}}{2})
    \Gamma_{n_2,n_1}^{\beta,\alpha} ({\bf k}- \frac{{\bf q}}{2},{\bf k}+ \frac{{\bf q}}{2})
   G_{n_1}({\bf k}- \frac{{\bf q}}{2},\omega_n -\frac{\Omega}{2})
   G_{n_2}({\bf k}+ \frac{{\bf q}}{2},\omega_n+\frac{\Omega}{2}),
\label{new3}
    \end{eqnarray}
\end{widetext}
where $G_i({\bf k}, \omega_n)$ is the Green's function of the
fermion in band $i$ at the Matsubara frequency, $\omega_n=\pi T
(2n +1)$. The tensor $\Pi_{ij}$ is diagonal, and its $zz$ ($xx$
and $yy$) component is the longitudinal (transverse) polarization.

The change in the static polarization upon entering the AFM state,
$\Pi (q \rightarrow 0,0)-\Pi_{m=0} (q \rightarrow
0,0)$ extends over all energies up to the fermionic bandwidth.
This is expected since this difference has to compensate negative
$\xi_0^{-2}$ that also
comes from high energy fermion physics. The relationship between
$m$ and $\xi_0^{-2}$ then  depends on the regularization chosen for the
high energies. We will simply measure the distance from the QCP on
the AFM side in terms of $m$, and use the Goldstone
condition $\xi^{-2}_0 + \Pi_{\perp} (q \rightarrow 0, 0) =0$. In contrast,
${\widetilde \Pi}_{ij}
(q, \Omega)$ are independent of the cutoff procedure and are fully
accounted for in the low-energy theory. It is essential that
 $\Pi_{\parallel} (q \rightarrow 0, 0) -
 \Pi_{\perp} (q \rightarrow 0, 0) = O(1/W)$, and hence to
this order, $\xi^{-2}_0 + \Pi_{\parallel} (q, \Omega)={\tilde \Pi}_{\parallel} (q, \Omega)$.

Evaluation of $\widetilde{\Pi}_\|$ and ${\widetilde \Pi}_\perp$ is
straightforward. We consider a two-dimensional Fermi surface with
tetragonal symmetry, so that there are four pairs of hot spots
which contribute additively to $\Pi (q, \Omega)$. Without loss  of
generality we choose the Fermi velocities at the hot spots ${\bf
k}_F$ and ${\bf k}_F +{\bf Q}$ along $x$ and $y$ directions,
respectively. It is convenient to introduce $k_\pm=(k_x\pm
k_y)/2$, so that $\epsilon_{\bf k}=v_F k_+$ and $\delta_{\bf
k}=v_F k_-$. These directions are inequivalent as the Fermi
surface exists for arbitrary ${\bf k}_-$, but only for $|{\bf
k}_+|> m/v_F$, see Fig.1. Evaluating the integrals in
Eq.(\ref{new3}) we find for a given pair of ``hot spots'' ($\Omega
>0$)
\begin{widetext}
    \begin{eqnarray}
    \label{Pi-perp}
   && {\widetilde \Pi}_{\perp}({\bf q},\Omega)=
    \frac{\gamma \Omega{\rm{sgn}}(v_F q_+ -\Omega)}{8}
    \biggl[{\frac{4m^2}{(v_F q_+
    -\Omega)^2-v_F^2 q_-^2}-1}\biggr]^{-1/2},   \\
   \label{Pi-par}
 &&{\widetilde \Pi}_{\parallel}({\bf q},\Omega)=
    \frac{\gamma\Omega {\rm{sgn}}(\Omega-v_F q_+)}{8}
    \biggl[\frac{4m^2}{(v_F q_+
    -\Omega)^2-v_F^2 q_-^2}-1\biggr]^{1/2},
    \end{eqnarray}
\end{widetext}
where $\gamma = g^2 \chi_0/(2\pi v_F^2)$, and $\Omega = \Omega + i
\delta_\Omega$. The full $\Pi (q, \Omega)$ are the sums over the
pairs of hot spots, i.e. the sum of contributions from the wave
vectors $(\pm q_+, \pm q_-)$ and $(\pm q_-, \pm q_+)$.

The behavior of the susceptibilities is particularly simple if we
set  $q_{+} =0$ in Eqs.(\ref{Pi-perp})-(\ref{Pi-par}). When
$\Omega, v_F q_- \gg m$, ${\widetilde \Pi}_{\perp} (q, \Omega) =
{\widetilde \Pi}_{\parallel} (q, \Omega) \sim i \gamma |\Omega|$,
as at the QCP with $z=2$. For $m \gg \Omega \gg v_F q_-$,
${\widetilde \Pi}_{\perp} \propto \Omega^2/m$, i.e.,
$\chi_{\perp}^{-1} \propto q_-^2 - (\gamma/m) \Omega^2$. In the
same regime ${\widetilde \Pi}_{\parallel}\approx 2\gamma m$, and
therefore longitudinal excitations have a gap. In the opposite
limit $m \gg v_F q \gg \Omega$, we find ${\widetilde \Pi}_{\perp}
\propto i |\Omega| (v_F q)/m$ and ${\widetilde \Pi}_{\parallel}
\propto i m \Omega/v_F q$. Consequently, the two spin
susceptibilities have the forms given in Eq. (\ref{new2}), i.e.,
spin excitations are overdamped and {\it gapless}, with $z=1$
($z=3$) for the transverse (longitudinal) channel.

\begin{figure}
    \epsfxsize=2.5in\epsfbox{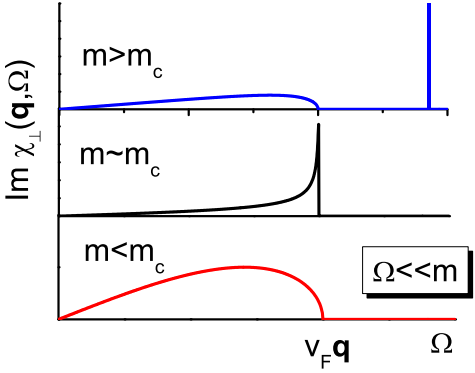}
    \caption{Transverse spin susceptibility
    at $\Omega, v_F q\ll m$. The propagating spin
    wave is manifested as a delta-function peak at $\Omega(q)$
    in the top panel.}
\label{fig2}
\end{figure}

That undamped spin waves exist only for $\Omega > v_F q$ means that
 propagating undamped spin waves exist only above
 a threshold value of staggered magnetization, $m_c \sim \gamma v^2_F \sim g$
(i.e., $\langle S^z_{\bf Q} \rangle = O(1)$). For $m<m_c$ the spin
wave would be at $\Omega< v_F q$, but it is damped there. At
$m=m_c$ the spin-wave pole first splits from the upper edge of the
low-frequency continuum and a propagating spin-wave emerges,
initially with the residue $Z \propto m - m_{min}$. This behavior
is illustrated in Fig. \ref{fig2}. For $\chi_{\parallel}$, there
is no distiction between different values of $m$ --
 in all cases, for $v_F q \ll m$,  there are two separate regions where
dissipation is present: one is at $\Omega < v_F q_-$, and the other
 at $\Omega > 2m - v_Fq_-$, as shown in
 Fig. \ref{fig3}.

\begin{figure}
    \epsfxsize=3in\epsfbox{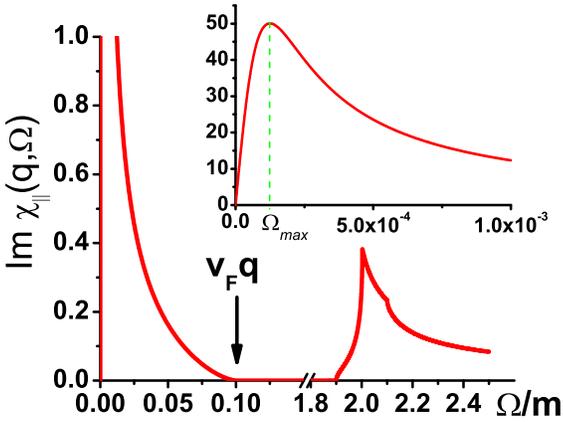}
    \caption{Dissipative part of the longitudinal susceptibility.
    Below energies $\sim 2m$ the onset of damping occurs at
    $\Omega=v_F q$. Notice the break in the horizontal scale.
     Inset: low energy behavior of ${\rm Im}\chi_\| ({\bf
    q}, \Omega)$. The maximum is at $\Omega_{max}\propto q^3$. Notice that
    in the Landau-damped region ${\rm Im}\chi_\| ({\bf
    q}, \Omega)$ is much larger than that in the continuum.}
\label{fig3}
\end{figure}

For a general ${\bf q}$, when both $q_+$ and $q_-$ are comparable,
the details of the momentum and frequency dependence of the
susceptibility are somewhat lengthly, but follow straightforwardly
from Eqs.(\ref{Pi-perp})-(\ref{Pi-par}). The Landau damping is now
effective for $\Omega < v_F (|q_+| +|q_-|)\ll 2m$. At small
frequencies $\chi_\|^{-1}\propto q^2-i \gamma m\Omega/(v_F
\sqrt{|q_+^2-q_-^2|})$.  The threshold for the spin-wave creation
along the direction $q_+=\alpha q_-$ is $m_c = \gamma v^2_F
(\alpha+1)^{3/2}(\sqrt \alpha +1)/(1+\alpha^2)$, and the spin-wave
pole first appears along the diagonals of the Brillouin Zone, at
$m\geq \gamma v^2_F$.

The regime $\Omega \ll v_F |q_+ -q_-| \ll m$
is particularly important for the electron
self-energy, $\Sigma({\bf k}, \omega)$. To the
lowest order in the interaction,
    \begin{eqnarray}
    \nonumber
    \Sigma_{n,\alpha}({\bf k}, \w)&=& T\sum_{{\bf
    q},\Omega \atop \beta n_1}
    \sigma^{(i)}_{\alpha\beta}\sigma^{(j)}_{\beta\alpha}
    \Gamma_{n n_1}^{\alpha\beta}({\bf k},{\bf q})
    \Gamma_{n_1 n}^{\beta\alpha}({\bf q},{\bf k})
    \\
    \label{Sigma}
    &&
    \times
    \chi_{ij}({\bf k-q}, \w-\Omega)
    G_{n_1}({\bf q},\Omega).
    \end{eqnarray}
We find that there are two non-trivial contributions to the self
energy. The first comes from the large momentum transfers $v_F
q\geq 2m$ and gives $\Sigma_1^{(1)}\approx (i g^2\sqrt{2}/v_F)
(\sqrt{4m^2-i\gamma a\omega}-2m)\propto \omega/m$ at small
energies.  At QCP this term gave rise to the $\sqrt{i\omega}$
dependence of the self energy at hot spots \cite{acs}. Finite
staggered magnetization, $m$, renders it harmless for the
fermions.

The second contribution appears for $m \neq 0$ from the
interaction between the electrons and the Landau damped
longitudinal spin excitations. This contribution is dominated  by
$v_F q \propto \Omega^{1/3}$ for which the criterium $v_F q >
\Omega$ is satisfied. In contrast to the transverse Goldstone mode
the interaction vertex for $\chi_\|$ is not reduced by Adler's
symmetry principle. As a result, the corresponding contribution to
the self-energy is $\Sigma_1^{2} \approx (\pi/4\sqrt 3) (g^2/v_F)
(i\omega)^{2/3}/(m\gamma)^{1/3}$, as in $z=3$ critical
 theories~\cite{qc3}. The fractional $\omega$ dependence with exponent
$2/3$ implies that both $Re \Sigma_1$ and $Im \Sigma_1$ are of the
same order, giving rise to a non-Fermi liquid behavior.
 Furthermore, as the Landau damping is due to small
momentum transfer, the NFL form of the self-energy persists
everywhere on the Fermi surface, not only at the former hot spots.
This is not surprising since, in the presence of the long-range
AFM order, the entire Fermi surface becomes ``hot'' as  the points
${\bf k}_F$ and ${\bf k}_F +{\bf Q}$ become equivalent. Finally,
in complete analogy to $z=3$ gauge and
ferromagnetic critical theories~\cite{qc3},  vertex
corrections and the momentum-dependent part of the self-energy are
small for $m \leq g \ll W$. Hence the result $\Sigma \propto (-i
\omega)^{2/3}$ remains valid to all orders in the interaction.

We believe that this scenario of Fermi liquid breakdown on the
ordered side of QCP is quite general. The requirement that the gap
in the optical mode of the order parameter dissolves into the
Landau damping due to small momentum scattering in the enlarged
unit cell is satisfied not only for AFM, but also for CDW and
other transitions with a finite ${\bf Q}$.

The most obvious experimentally observable consequence of this
scenario  is the existence of the two regions of $\Omega$
 where $Im \chi_{\parallel} (q, \Omega) \neq 0$ (see Fig. \ref{fig3}).
 This can be probed by inelastic neutron scattering experiments.
 The fractional exponent in the self energy also
leads to the non-Fermi liquid, quantum-critical
 behavior for the temperature
dependence of the electronic specific heat and
resistivity~\cite{millis}, which is accessible experimentally.

We are grateful to Aspen Center for Physics for its hospitality.
This work was supported by the US DOE (I.~V.) and by NSF DMR
0240238 (A.~C.). We thank E. Abrahams, I. Aleiner, Ar. Abanov, D. Morr,
D. Pines, D. Scalapino,  and J. Schmalian for useful conversations.

\end{document}